\documentclass{article}

\usepackage{arxiv}

\usepackage{graphicx}
\usepackage{tikz-qtree}
\usepackage{wasysym}
\usepackage{caption}
\usepackage{listings}
\usepackage{fancybox}
\usepackage{color}
\usepackage{lstlinebgrd}
\usepackage{xcolor}
\usepackage{parcolumns}
\usepackage{array}
\newcolumntype{P}[1]{>{\centering\arraybackslash}p{#1}}
\makeatletter
\newenvironment{CenteredBox}{%
\begin{Sbox}}{
\end{Sbox}\centerline{\parbox{\wd\@Sbox}{\TheSbox}}}
\makeatother

\title{Automatic Repair and Type Binding of Undeclared Variables using Neural Networks}

\author{
  Venkatesh Theru Mohan \\
  Department of Computer Science\\
  Iowa State University\\
  \texttt{venkytm@iastate.edu} \\
   \And
  Ali Jannesari \\
  Department of Computer Science\\
  Iowa State University\\
  \texttt{jannesar@iastate.edu} \\
}

\begin{document}
\maketitle

\begin{abstract}
{Deep learning had been used in program analysis for the prediction of hidden software defects using software defect datasets, security vulnerabilities using generative adversarial networks as well as identifying syntax errors by learning a trained neural machine translation on program codes. However, all these approaches either require defect datasets or bug-free source codes that are executable for training the deep learning model. Our neural network model is neither trained with any defect datasets nor bug-free programming source codes, instead it is trained using structural semantic details of Abstract Syntax Tree (AST) where each node represents a construct appearing in the source code. This model is implemented to fix one of the most common semantic errors, such as undeclared variable errors as well as infer their type information before program compilation. By this approach, the model has achieved in correctly locating and identifying 81$\%$ of the programs on prutor dataset of 1059 programs with only undeclared variable errors and also inferring their types correctly in 80$\%$ of the programs.}
\end{abstract}


\section{Introduction}
In the recent decades, there had been some significant contributions in the automation of program analysis tasks. Neural networks had been frequently used either for detection or generation tasks in programming language processing similar to natural language processing. It had been employed in detecting software defects, as well as prediction of errors by using software metrics \cite{jayanthi2018software}. 
\newline
\\
There have been some state-of-art performances achieved by generative adversarial networks and neural machine translation systems on language translation tasks in Natural Language Processing (NLP) that led to the deployment of these systems on error correction tasks in the programming source codes. Neural machine Translation (NMT) in \cite{ahmed2018compilation} builds and trains a single neural network model using a labelled set of paired examples that results in translation from the input directly. They are end-to-end in the sense that they have the ability to learn the source text directly by mapping them to the corresponding target text and uses an encoder-decoder(usually made up of Recurrent Neural Networks (RNN) units) approach for applying the transformations where encoder consists of sequences of source text and output consists of sequences of target text. Generative Adversarial Networks (GAN) trains a neural network model to predict security vulnerabilities in \cite{harer2018learning} without the necessity of being a paired set of source domain containing buggy codes and target domain consisiting of bug-free codes, instead being a bijection mapping. Generally, in this labelled set of paired examples or unpaired examples, the neural network model is trained on the set of positive examples where the mapping takes place between target sequences, made up of positive examples that are bug-free and compiling without any errors and source sequences consisting of negative examples that contains bugs within the code making the compilation to fail. In sequence-to-sequence learning systems such as NMT, it is flexible to train and learn the model with a labelled set of paired examples, but consider the scenario where there are no paired examples, further to put it in a simpler context, consider a real-case scenario where some common syntax or semantic errors are being committed by freshers or novice programmers working in a software industry or student submissions of programming assignments in coding competitions and there are no positive or bug-free reference examples, then learning becomes difficult for neural network approaches and neural machine translation models.
\\
\\
In our model, there are no positive examples to train and focus is only on the negative buggy examples specifically, the common semantic error caused due to undeclared variables which is often committed and unperceived by the novice programmers. The model is instead trained based on the structural and semantic elements, that is the non-terminal and terminal nodes of the programming source codes captured from the abstract syntax tree representation. The type of the undeclared variables is also inferred by performing type binding using the semantic elements of AST representation that provides about the type information of the variables thereby saving the compiler's time in performing the type binding of those undeclared variables. The comprehensive information of the ASTs, the motive of Long Short-Term Memory (LSTM), detailed view of the training approach and implementation, the generation approach and different scenarios where undeclared variables will be caused and possible cases of type inference of them will be discussed in the upcoming sections of the paper.         
\section{Related Work}
 AST are the static intermediate model of a programming source code as discussed in \cite{4299919} where the compiler's analytic front-end parses it, constructs the AST model eventually passing it to the compiler's synthetic back-end to produce an assembly code for a specific machine and also used for program analysis/transformation. Often low-level concrete syntax tree representations have a complex structure and it is difficult to characterize the semantics especially in the poorly understood domains, so in \cite{wile1997abstract}, describes a transformation process to get a good abstract syntax representation from low-level concrete specification where modern tools rely on its ability to analyze, simulate and synthesize programs easily in language processing. 
 \newline
 \\
 In the past recent decades, deep learning had achieved various scales in text domain such as natural language especially in neural machine translation tasks and used it successfully even for programming language processing tasks as well. Some of the recent works that had achieved empiricial success on neural machine translation include dialogue response generation, summarization and text generation tasks as explained in detail in \cite{kosovan2017dialogue}, also NMT tasks are very much useful in translation from one language to another like in \cite{choudhary2018neural} where NMT encoder-decoder model had been implemented using word embedding along with byte-pair encoding method to develop an efficent translation mechanism from English-Tamil. The performance of such text generation tasks such as NMT is enhanced with attention mechanism in \cite{bahdanau2014neural} to automatically search for a context of a source text that is relevant for prediction of target words and an ensemble model of global and local attention mechanism of \cite{luong2015effective} improving the performance. 
 \\
 \\
 Natural language generation having a discrete output space had also been implemented by generative adversarial networks (introduced by \cite{goodfellow2014generative}) on generating sentences from context-free grammars and probabilistic grammars as shown in \cite{rajeswar2017adversarial}. The quality of the text generation had been improvised by conditioning information on generated samples to produce realistic natural looking sentences compared to maximum likelihood training procedures. Text generation has also shown some encouraging results in \cite{guo2018long} in which the discriminator of the GAN leaks its own high-level features to the generator at each of the generator steps thus making the scalar guiding signal available continuously throughout the generative process. 
 \\
 \\
 In \cite{liu2016neural}, prediction of next tokens in the source code is implemented using LSTM neural networks where the model trains and learns associated subsequent nodes for code completion, given a partial AST containing left subset of nodes or semantic features with respect to a subtree. The efficiency of AST in extracting tokens and comparing the source codes based on them and the use of deep learning in classifying duplicate/clone codes can be helpful in software code maintenance as seen in \cite{li2017cclearner} where maintaining duplicate codes for reuse in order to improve productivity of programming becomes a burden when there are inconsistencies caused due to bug fixes and program modifications in the original code at multiple locations.
 \\
 \\
 The significance of AST representation of source code can further be noted in \cite{dam2017automatic} where LSTM neural networks are leveraged to capture the long contextual relationships between semantic features to identify related code elements in order to predict the software vulnerabilities which causes a security threat or makes the program buggy. GAN approach is used in \cite{harer2018learning} for repairing vulnerabilities in source codes without any paired examples or bijections by mapping from non-buggy source domain to buggy target domain and training the discriminator using the loss that is generated between real examples and NMT-generated outputs from generator of desired output.
 \\
 Syntax errors poses a threat as it fails the compilation and some recent techniques such as \cite{ahmed2018compilation} where a RNN model is learnt on syntactically correct, executing student programming course submissions to model all valid sequences of token and use a prefix token sequence which is from the beginning of the program till the error location and is used to predict the following sequence that are able to automate the repair of errors present in corresponding locations of code. Sequence-to-sequence NMT with attention mechanism is learned iteratively in repairing syntax errors in \cite{gupta2017deepfix} using the tokenized vector representation of the program and is used to predict the erroneous program locations and the fixing statement without using any external compiler tools or any AST representation. A real-time feedback is given to the students enrolled in beginner level programming assignments in \cite{bhatia2016automated} of the compile-time syntax errors that are made by using RNN to predict the target lines from syntactically correct submissions given the source error lines from wrong submissions and a abstract version of top ranked suggestion of error fix is presented as feedback. 
\begin{figure}
\begin{CenteredBox}
\begin{lstlisting}[xleftmargin=.1\textwidth,linebackgroundcolor={%
        \ifnum\value{lstnumber}=7
                \color{red!40}
            \fi},linebackgroundwidth=18em,numbersep=25pt, basicstyle=\small]
<@\textcolor{blue}{\#include}@> <stdio.h>
<@\textcolor{blue}{int}@> main()
{
    <@\textcolor{blue}{int}@> i,max,j,n,m,y;
    scanf("%d %d",&n,&m);
    <@\textcolor{blue}{for}@>(i=1;i<=n;i++){
      s=0;
      <@\textcolor{blue}{for}@>(j=1;j<=m;j++){
      scanf("%d",&y);    
      s=s+j;
      }
      <@\textcolor{blue}{if}@>(max<s)
      max=s;
    }
    <@\textcolor{blue}{return}@> 0;
}
\end{lstlisting}
\end{CenteredBox}
\caption{An example illustrating the undeclared variable "s" that is frequently used in the program is caught by the compiler}
\end{figure}
\noindent
\section{Approach}
This section covers some examples of undeclared variables and also the importance of semantic analysis to determine the type information of those undeclared variables. We introduce the Abstract Syntax Trees (AST) that is used as the input, deployment of the LSTM RNN for training the deep learning model, semantic analysis determining the types of undeclared variables, and the generation approach by performing the serialization/deserialization of the AST in order to get back the clean and bug-free source code.  
\subsection{Motivating Examples}
The most frequent semantic error that goes unnoticeable by novice programmers is the undeclared variables. The cause of this error is due to the variables being undeclared or another common cause will be usually through spelling mistakes which makes it the first occurrence in the program. 
\begin{figure}
\begin{CenteredBox}
\begin{lstlisting}[xleftmargin=.1\textwidth,linebackgroundcolor={%
        \ifnum\value{lstnumber}=16
                \color{red!40}
            \fi},linebackgroundwidth=19.5em,numbersep=18pt, basicstyle=\small]
<@\textcolor{blue}{\#include}@> <stdio.h>
<@\textcolor{blue}{int}@> main()
{
    <@\textcolor{blue}{int}@> n,m;
    <@\textcolor{blue}{int}@> i,j;
    <@\textcolor{blue}{int}@> a[20];
    <@\textcolor{blue}{int}@> sum=0;
    scanf("%d%d",&n,&m);
    <@\textcolor{blue}{for}@>(i=0;i<n;i++){
        <@\textcolor{blue}{for}@>(j=0;j<m;j++){
           scanf("%d",&a[j]);
           sum=sum+a[j];
        }
        printf("%d\n",sum);
        i++;
        J++;
    }
    <@\textcolor{blue}{return}@> 0;
}
\end{lstlisting}
\end{CenteredBox}
\caption{An example of compiler error caused by an undeclared variable "J" that is used once in the program}
\end{figure}
The main challenge lies in the fact in determining whether the variable is an identifier, arrays, pointers or pointers-to-pointers and also in concluding about the type of the variable if it is an integer, float, character, double, long int and so on. The C99 standard\footnote{www.open-std.org/jtc1/sc22/wg14/www/docs/n1256.pdf}, removes the implicit integer rule that states a variable declared without an explicit data type is assumed to be integer which was previously defined in C89 standard\footnote{https://www.pdf-archive.com/2014/10/02/ansi-iso-9899-1990-1/ansi-iso-9899-1990-1.pdf}. Therefore, there is a need for determining the variables that are undeclared along with its type, else the compiler will throw an error. 
\\
\subsection{Abstract Syntax Trees}
Generally, any programming language whether statically or dynamically typed follows an unambiguous context-free grammar language where there exists not more than one leftmost derivations or rightmost derivations of non-terminals, or more precisely there is always a unique parse tree for each string of the language generated by it.
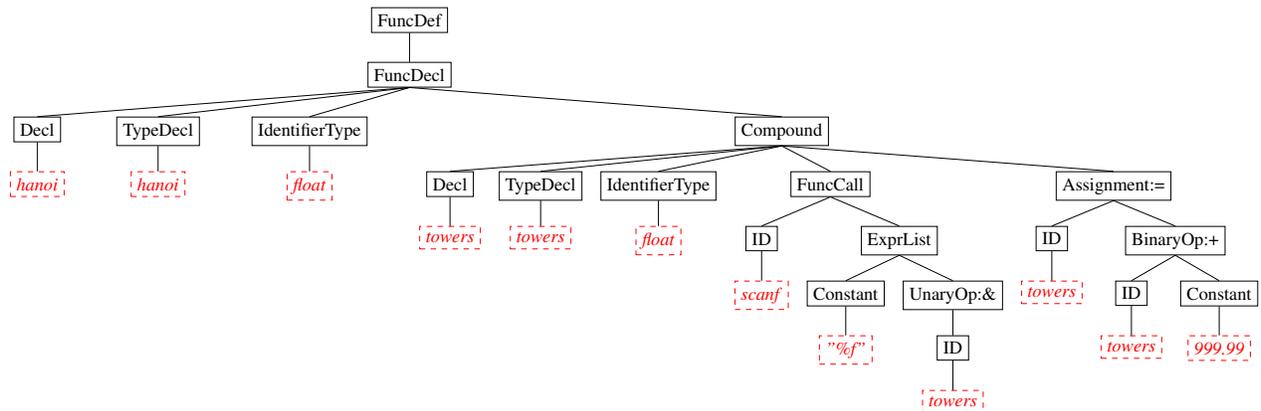
\begin{figure*}
\begin{tikzpicture}[scale=0.69,sibling distance=10pt]
\tikzset{every tree node/.style={align=center,anchor=north}}
\tikzset{level 1/.style={sibling distance=30mm},level 2/.style={sibling distance=10mm}}
\Tree [.\node[draw,color=black]{FuncDef}; [.\node[draw,color=black] {FuncDecl}; [.\node[draw,color=black]{Decl}; \node[draw,dashed,color=red]{\textit{hanoi}}; ] [.\node[draw,color=black] {TypeDecl}; \node[draw,dashed,color=red]{\textit{hanoi}}; ] [.\node[draw,color=black] {IdentifierType}; \node[draw,dashed,color=red]{\textit{float}}; ] [.\node[draw,color=black] {Compound}; [.\node[draw,color=black] {Decl}; \node[draw,dashed,color=red]{\textit{towers}}; ] [.\node[draw,color=black] {TypeDecl}; \node[draw,dashed,color=red]{\textit{towers}}; ] [.\node[draw,color=black] {IdentifierType}; \node[draw,dashed,color=red]{\textit{float}}; ] [.\node[draw,color=black] {FuncCall}; [.\node[draw,color=black] {ID}; \node[draw,dashed,color=red]{\textit{scanf}}; ] [.\node[draw,color=black] {ExprList}; [.\node[draw,color=black]{Constant}; \node[draw,dashed,color=red]{\textit{"\%f"}}; ] [.\node[draw,color=black]{UnaryOp:\&}; [.\node[draw,color=black]{ID}; \node[draw,dashed,color=red]{\textit{towers}}; ]] ] ]  [.\node[draw,color=black] {Assignment:=}; [.\node[draw,color=black] {ID}; \node[draw,dashed,color=red]{\textit{towers}}; ] [.\node[draw,color=black] {BinaryOp:+}; [.\node[draw,color=black] {ID}; \node[draw,dashed,color=red]{\textit{towers}}; ] [.\node[draw,color=black] {Constant}; \node[draw,dashed,color=red]{\textit{999.99}}; ] ] ] ] ] ]
\end{tikzpicture}
\caption{An example illustration of Abstract Syntax Trees(AST) of C program with non-terminals at the root as well as internal nodes and terminals at the leaf nodes of the tree.}
\end{figure*}
 Parse trees are formed from a concrete context-free grammar and are not suitable for performing syntax or semantic analysis due to their complex representation. Nevertheless, the Abstract Syntax Tree are the derivation trees following an abstract grammar is used as an input for syntax/semantic analysis at compile-time. It is a rooted tree representation of an abstract syntactical structure of a programming language construct where the non-terminals represents non-leaf nodes and the terminals forms the children nodes. Some of the non-terminals in C language are \texttt{Decl, IdentifierType, For, TypeDecl, ExprList, FuncDef, ArrayDecl} etc. Terminal nodes includes any string literals, numerical literals, variable names, operators, keywords etc. Figure 3 shows an example AST representation where a node enclosed in a rectangle box and black-colored depicts the non-terminal nodes (eg.,\texttt{FuncDef}), node that are enclosed in dashed box and red-colored depicts the terminal nodes or tokens of the source code.
\subsection{Model}
This subsection covers the basics of LSTM-RNN and the prediction model that is subsequently used for generation approach. \\ \\
Long Short Term Memory(LSTM) recurrent neural networks have special memory units in the form of self-loops to produce paths so that information can be maintained for longer durations of time. LSTM is preferred over vanilla RNN due to the fact that the former tends to avoid vanishing or exploding gradient problem that occurs when trying to learn long term dependencies and store it in memory cells during backpropagation. This occurs when many deep layers with specific activation functions like sigmoid are used for training, it smoothens a region of input space into an output space between 0 and 1, then even a high change in input region effects almost negligible change in output region, thereby making the gradients/error signals of a long-term interaction becomes vanishingly very small. Further, the vanilla RNNs are affected by information morphology problem in which information contained at a prior state is lost due to non-linearites between the input and output space. LSTM avoids this problem by ensuring a constant unit activation function and uses gates to control the information flow between the memory cell and the outside layers without any inference. LSTM uses three gates namely forget gate, input gate and output gate layers. The forget gate layer decides the information that is needed to be stored or erased from the LSTM cell state where the decision is made by the sigmoid layer outputting a number between 0 to 1.
\begin{equation}
    f\textsubscript{i}\textsuperscript{(t)} = \sigma(b\textsubscript{i}\textsuperscript{f} + \sum\limits_{j} U\textsubscript{i,j}\textsuperscript{f}x\textsubscript{j}\textsuperscript{(t)} + \sum\limits_{j} W\textsubscript{i,j}\textsuperscript{f}h\textsubscript{j}\textsuperscript{(t-1)} )
\end{equation}
where x\textsuperscript{(t)} is the input vector at current timestep t and h\textsuperscript{(t)} is the current hidden layer vector at timestep t, and b\textsuperscript{f}, U\textsuperscript{f}, W\textsuperscript{f} are bias units, input weights and recurrent weights of forget gate units f\textsubscript{i}\textsuperscript{(t)}. 
\\
The input gate layer controls the flow of new information that is being stored in the LSTM cell state s\textsubscript{i}\textsuperscript{(t)} conditioned with a self-loop weight f\textsubscript{i}\textsuperscript{(t)}.
\begin{equation}
    g\textsubscript{i}\textsuperscript{(t)} = \sigma(b\textsubscript{i}\textsuperscript{g} + \sum\limits_{j} U\textsubscript{i,j}\textsuperscript{g}x\textsubscript{j}\textsuperscript{(t)} + \sum\limits_{j} W\textsubscript{i,j}\textsuperscript{g}h\textsubscript{j}\textsuperscript{(t-1)} )
\end{equation}
where x\textsuperscript{(t)} is input vector at current timestep t and h\textsuperscript{(t)} is current hidden layer vector at timestep t, and b\textsuperscript{g}, U\textsuperscript{g}, W\textsuperscript{g} are bias units, input weights and recurrent weights of input gate units g\textsubscript{i}\textsuperscript{(t)}.
\begin{equation}
    s\textsubscript{i}\textsuperscript{(t)} = f\textsubscript{i}\textsuperscript{(t)}s\textsubscript{i}\textsuperscript{(t-1)} + g\textsubscript{i}\textsuperscript{(t)} \sigma(b\textsubscript{i} + \sum\limits_{j} U\textsubscript{i,j}x\textsubscript{j}\textsuperscript{(t)} + \\  \sum\limits_{j} W\textsubscript{i,j}h\textsubscript{j}\textsuperscript{(t-1)} )
\end{equation}
where the parameters W,U and b represents the recurrent weights, input weights and bias units present in a LSTM cell. The output gate layer in the memory cell decides the pieces of information that is going to be output by the LSTM cell state. This is done by passing the cell state through a tanh layer and eventually multiplying by the sigmoid of the output gate. 
\begin{equation}
    q\textsubscript{i}\textsuperscript{(t)} = \sigma(b\textsubscript{i}\textsuperscript{o} + \sum\limits_{j} U\textsubscript{i,j}\textsuperscript{o}x\textsubscript{j}\textsuperscript{(t)} + \sum\limits_{j} W\textsubscript{i,j}\textsuperscript{o}h\textsubscript{j}\textsuperscript{(t-1)} )
\end{equation}
where b\textsuperscript{o},U\textsuperscript{o},W\textsuperscript{o} are the parametric units of the output gate q\textsubscript{i}\textsuperscript{(t)} that represents bias units, input weights and recurrent weights.
The output hidden state h\textsubscript{i}\textsuperscript{(t)} is obtained from output gate q\textsubscript{i}\textsuperscript{(t)} as follows:  
\begin{equation}
     h\textsubscript{i}\textsuperscript{(t)}= tanh(s\textsubscript{i}\textsuperscript{(t)})q\textsubscript{i}\textsuperscript{(t)}
\end{equation}
In our prediction model, we use non-terminals and terminals obtained from the AST as the input. The main ideology behind our model is to specify a declaration for any identifier that is used throughout a C program. Here, we assume identifier in our context that excludes  keywords and only includes alphanumeric variables. This in turn solves the complex problem of automatically fixing the undeclared variables.\\
\subsubsection{Declaration Classification and Prediction}
\texttt{ID} is the non-terminal node of the AST that represents an identifier excluding keywords as mentioned above. \texttt{Decl} is the non-terminal node of the AST representation that is entitled to represent the declaration of the identifier where its terminal node is the corresponding identifier itself. Similarly, \texttt{TypeDecl} and \texttt{IdentifierType} are the semantic elements that is used to represent the type specifier information of the identifier where identifier and its type are its corresponding terminals respectively.
\\
\\
For the classification purpose, the pair of non-terminal node \texttt{ID} and any terminal alphanumeric identifiers usually variables is augmented along with pairs of \texttt{Decl} and the respective identifier, \texttt{TypeDecl} and the identifier, \texttt{IdentifierType} and a generalized "type" referring to the corresponding types of those identifier variables so that they can be used for backsubstitution which will be explained later in the generation approach. After classification, the LSTM model is used to predict the \texttt{Decl}, \texttt{TypeDecl} and \texttt{IdentifierType} information for any alphanumeric identifier variable occurring with its corresponding non-terminal node \texttt{ID}. The classification model is sequential where the embedding layer, LSTM layer and softmax layers are stacked on top of each other sequentially as shown in Figure 4 are detailed below in this section.   \\  
\subsubsection{Embedding Layer of Non-Terminal and Terminal Nodes}
In our model, we do not use any pre-trained embeddings, instead it is trained simultaneously with the model. The sequences of input tokens are a combination of non-terminal and terminal nodes where in our model, we consider only 4 non-terminals \texttt{ID, Decl, TypeDecl, IdentifierType} and all the alphanumeric identifier variables as the terminals. These input tokens are formed by the concatenation of individual string encodings of non-terminal and terminal node vocabularies (that is discussed in evaluation section), subsequently embedding is computed on the integer encodings(converted from string) for performing the training of the model. The embeddings are computed as follows:
\begin{equation}
    E\textsubscript{i}= A*concat(N\textsubscript{i}T\textsubscript{i})
\end{equation}
where A is $K\times V\textsubscript{N,T}$ matrix where K is the embedding vector size and V\textsubscript{N,T} is the vocabulary size formed by the concatenated encodings of non-terminal and terminal nodes.
\\
\subsubsection{LSTM Layer}
The sequences of embedded tokens are passed on to the LSTM layer containing LSTM memory cells where each cell state stores information of the previous state and are controlled by the forget gate, input gate and output gate layers as mentioned above in equations (1),(2),(3),(4). Each LSTM cell state takes inputs from its previous LSTM cell hidden state
\begin{figure*}
\begin{center}
\includegraphics[width=320pt,height=220pt]{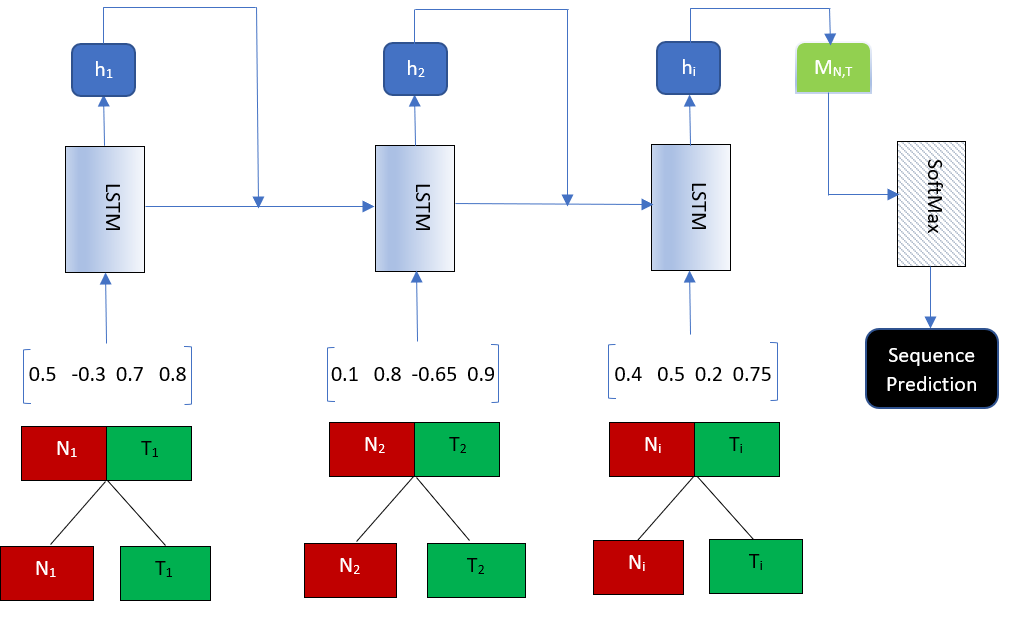}
\caption{Illustration of approach showing concatenation of non-terminal and terminal node embeddings extracted from AST being used as the inputs to LSTM model for the sequence classification and prediction.}
\end{center}
\end{figure*}
h\textsubscript{i-1}, state information s\textsubscript{i} as well as the input tokens and outputs the hidden state h\textsubscript{i} of LSTM cell as in equation (5) where the LSTM layers can be seen from Figure 4. 
\\
\subsubsection{Dense Softmax Activation Layer}
The last LSTM memory cell state's output hidden state of the LSTM layer is passed to the softmax activation layer to predict the sequences of non-terminals \texttt{Decl}, \texttt{TypeDecl} and \texttt{IdentifierType} given the non-terminal \texttt{ID}. The predicted output sequences at timestep t (or fixed input sequence length) are represented by \begin{math}\hat{y}(t)\end{math} and is formulated as:
\begin{equation}
     \hat{y}(t) = softmax(b\textsubscript{N,T}+ M\textsubscript{N,T}h\textsubscript{(t)})
\end{equation}
where b\textsubscript{N,T} is the bias unit of softmax layer with size of V\textsubscript{N,T} dimensional vector, M\textsubscript{N,T} is a weight matrix of size $K \times V\textsubscript{N,T}$ and h\textsubscript{(t)} is the hidden state of the LSTM cells at each timestep t. 
\section{Evaluation and Results} 
\subsection{Dataset}
The dataset that used in this approach is prutor\footnote{https://www.cse.iitk.ac.in/users/karkare/prutor/prutor-deepfix-09-12-2017.zip} which is a database that has student coding submissions for university programming assignments. It contains a set of 53478 C programs out of which there are 6978 erroneous programs which contains multiple and single line syntax as well as semantic errors. Out of 6978 programs, 1059 programs contains only undeclared variable errors which is the main focus of our evaluation. 
\subsection{Preprocessing and Training Details} 
We use pycparser\footnote{https://pypi.org/project/pycparser/}, that acts as a front-end of the C compiler to parse source code of C language in python. AST are obtained as an output for the source code after the parsing stage and are stored in text files. 
\newline 
\\
The source code is preprocessed in the form of tokens that represents the terminal nodes of its corresponding AST. Since the set of tokens are uncontinuous, discrete and in its textual form, it must be encoded into sequences of numerical vectors to be used for training the model. Additionally, there are 47 fixed set of non-terminals in C language that are encoded as in Figure 5. The terminals can be keywords, strings, data types, integers or floating point numbers. The terminals of the above mentioned categories are encoded separately in a specific range of numbers. Now, the data for training is prepared by concatenating the encodings of non-terminals and terminals together. The individual encodings in the form of integers are converted to strings initially and after concatenation, they are converted back to integers. For example, the non-terminal \texttt{IdentifierType} is encoded as 9 in the dictionary of non-terminals and converted to '9', if the terminals are data types like int, float, long etc. then they are encoded as 111111 referring to a generalized 'type' and converted to '111111'. Therefore, the concatenation of the non-terminals and terminals are mapped accordingly and stored in separate vocabulary. This vocabulary set is used in performing the training of the model. One-hot encoding approach is used to perform categorical multi-class classification to represent the elements of vocabulary as vectors with each of them of vocabulary size containing 1 at the corresponding index and rest of them are 0.   
\newline 
\\
\textbf{Training Details:}\\ 
The training experiment is performed by using embedding dimension of size 512 and two LSTM layers are used each with number of hidden units as 512 and a dropout of 0.5. The input sequence length is 1, batch size is 3, vocabulary size is 583 and the total number of sequences is 2319. The dense layer is used for forming a fully connected layer in which each of the input layer nodes is connected with every hidden and output layer nodes. The activation function used in our model is softmax function because of its efficiency in dealing with multi-class classification problems compared to sigmoid and ReLU due to the fact that the outputs of the softmax is a categorical probability distribution summing to 1 and lying between 0 to 1. The total number of units of Dense layer is equal to the vocabulary size. The vocabulary formed from concatenation is split into training and testing data where test data size is 0.2 and the split rule percentage used is 80/20. The loss function used is categorical cross-entropy function. The optimizer used is RMSprop with a learning rate of 0.01 as it is better in handling non-local extremum points and has a constant initial global learning rate compared to optimizers such as Stochastic gradient descent optimizer. 
\begin{figure}
\includegraphics[width=\linewidth,height=120pt]{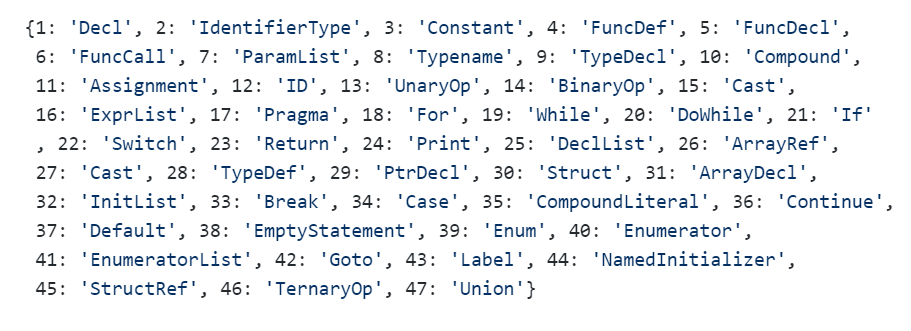}
\caption{The vocabulary of all non-terminal nodes of AST for C programs}
\end{figure}
\subsection{Generation Approach}
During generation, one-hot encoding is used to represent the new unseen sequences of the nonterminal \texttt{ID} and a terminal variable as a categorical distribution and the output class is classified as the sequences of non-terminals \texttt{Decl, TypeDecl, IdentifierType} along with the corresponding terminal variables.
\subsubsection{AST Transformation}
The program fix approach is carried out through an AST transformation performed by augmenting the predicted output sequences in each of the program's AST syntactical structure for the terminal variables associated with its corresponding non-terminal node \texttt{ID}. This augmentation is carried out on each of the source code by performing a check on declaration of the variables in the vocabulary set of concatenated encodings of non-terminal and terminal nodes as created previously using the predicted output sequence of the non-terminal \texttt{Decl} and the associated terminal variable. If any of the predicted output sequences does not match with the declared variables present in a source code, then the output sequence \texttt{Decl} containing the particular terminal variable is augmented to the original AST structure of the code through serialization and deserialization that will be described below.  
\begin{figure*}
\begin{lstlisting}[xleftmargin=.3\textwidth, xrightmargin=.3\textwidth, basicstyle=\small,numbers=none]
{
    "_nodetype": "Decl",
    "bitsize": null,
    "funcspec": [],
    "init": null,
    "name": "j",
    "quals": [],
    "storage": [],
    "type": {
       "_nodetype": "TypeDecl",
       "declname": "j",
       "quals": [],
       "type": {
          "_nodetype": "IdentifierType",
          "names": [
              "int"
           ]
       }
   }
}
\end{lstlisting}
\caption{Example demo of JSON object containing Decl, TypeDecl and IdentifierType nodes.}
\end{figure*}
\noindent
\\
\subsubsection{Serialization and Deserialization}
Serialization is implemented using pycparser by transforming data structures such as nodes of python Node object by traversing the AST (the nodes being obtained by parsing the source code from pycparser) recursively into a dictionary object representation which is then subsequently serialized into a JSON object that is understood by Pycparser in deserializing it back to a dictionary object and thereby consequently back to AST Node objects. An example JSON object representation of a AST node object is shown in Figure 6 where the \texttt{\_nodetype} key refers to the different types of non-terminal nodes such as \texttt{Decl, ArrayDecl, TypeDecl, IdentifierType}. The \texttt{TypeDecl} and \texttt{ArrayDecl} are the child nodes of \texttt{Decl} and \texttt{IdentifierType} is the child node of the intermediate non-terminal \texttt{TypeDecl} node. The key  \texttt{name} refers to the terminal variables, \texttt{type} refers to the datatypes and \texttt{coord} refers to the \texttt{Coord} node of the AST indicating the location of the object in the corresponding source code.    
\\
\\
The serialization and deserialization is carried out when there is an AST transformation performed as mentioned above, so that the transformation is taking place consistently without disintegrating the program thereby ensuring maximal quality of the source code.
\subsubsection{Pre-Compile Time Type Binding and Analysis Results}
After determining and performing the augmentation of the undeclared terminal variables, the type of those undeclared variables is determined before compiling the program and is augmented to the original AST structure. The type binding is performed by determining the type of a \texttt{lvalue} from its \texttt{rvalue} in an assignment statement or finding the type of a undeclared variable from its neighboring variables in an expression of a program. 
\\
\\
The type of the undeclared variables is determined and drawn from the following cases:
\\
\\
\textbf{Case 1:} When the \texttt{rvalue} is a constant, where the non-terminal is \texttt{Constant} and is of integer type, then the \texttt{lvalue} whose non-terminal is \texttt{ID}, is assigned \texttt{integer} as seen from the Figure 7, in the assignment statement \textbf{i=0} on left side of figure, "i" is undeclared and is assigned \texttt{integer}.
\\
\\
\begin{minipage}{.4\textwidth}
\begin{lstlisting}[xleftmargin=.2\textwidth,linebackgroundcolor={%
        \ifnum\value{lstnumber}=6
                \color{red!40}
            \fi},linebackgroundwidth=14em,numbersep=10pt,basicstyle=\small]
<@\textcolor{blue}{int}@> main()
{
<@\textcolor{blue}{int}@> k,n,x,a[100];
scanf("%d", &k);
scanf("%d", &n);
<@\textcolor{blue}{for}@>(i=0;i<n;i++)
   scanf("%d", &a[i]);
<@\textcolor{blue}{return}@> 0;
}

\end{lstlisting}
\end{minipage}\hfill\vline\hfill
\begin{minipage}{.8\textwidth}
\begin{lstlisting}[xrightmargin=.5\textwidth, linebackgroundcolor={%
        \ifnum\value{lstnumber}=3
                \color{green!40}
            \fi},linebackgroundwidth=14em,numbersep=10pt,basicstyle=\small]
<@\textcolor{blue}{int}@> main()
{
<@\textcolor{blue}{int}@> i;
<@\textcolor{blue}{int}@> k;
<@\textcolor{blue}{int}@> n;
<@\textcolor{blue}{int}@> x;
<@\textcolor{blue}{int}@> a[100];
scanf("%d", &k);
scanf("%d", &n);
<@\textcolor{blue}{for}@> (i = 0; i < n; i++)
    scanf("%d", &a[i]);
<@\textcolor{blue}{return}@> 0;
}

\end{lstlisting}
\end{minipage}
\captionof{figure}{Case 1 demonstrating location of error in the for loop statement involving undeclared identifier "i" and the fix of it}
\noindent
\\
\textbf{Case 2:} In this case, if the \texttt{rvalue} is an identifier that refers to an array element whose non-terminal is \texttt{ID} and its non-terminal parent is \texttt{ArrayRef}, then the \texttt{lvalue} terminal variable with non-terminal \texttt{ID} is assigned to the respective type of \texttt{rvalue} element. We can see in Figure 8 where "b" is undeclared and is assigned to \texttt{integer} type from array "n[1000]" in the statement \textbf{b=n[i]}.
\\
\\
\begin{minipage}{.4\textwidth}
\begin{lstlisting}[xleftmargin=.2\textwidth,linebackgroundcolor={%
        \ifnum\value{lstnumber}=13
                \color{red!40}
            \fi},linebackgroundwidth=14em,numbersep=10pt,basicstyle=\small]
<@\textcolor{blue}{int}@> main()
{
<@\textcolor{blue}{int}@> n[1000],a[500],nm,i,j,ln,flag=0;
scanf("%d\n",&ln);
scanf("%d\n",&nm);
<@\textcolor{blue}{for}@>(i=0;i<500;i++)
{
    a[i]=0;
}
<@\textcolor{blue}{for}@>(i=0;i<nm;i++)
{
    scanf("%d ",&nm);
    c=n[i];
}
<@\textcolor{blue}{return}@> 0;
}
\end{lstlisting}
\end{minipage}\hfill\vline\hfill
\begin{minipage}{.8\textwidth}
\begin{lstlisting}[xrightmargin=.5\textwidth, linebackgroundcolor={%
        \ifnum\value{lstnumber}=3
                \color{green!40}
            \fi},linebackgroundwidth=14.5em,numbersep=10pt,basicstyle=\small]
<@\textcolor{blue}{int}@> main()
{
<@\textcolor{blue}{int}@> c;
<@\textcolor{blue}{int}@> n[1000];
<@\textcolor{blue}{int}@> a[500];
<@\textcolor{blue}{int}@> nm;
<@\textcolor{blue}{int}@> i;
<@\textcolor{blue}{int}@> j;
<@\textcolor{blue}{int}@> ln;
<@\textcolor{blue}{int}@> flag = 0;
scanf("%d""\n", &ln);
scanf("%d""\n", &nm);
<@\textcolor{blue}{for}@> (i = 0; i < 500; i++)
{
    a[i] = 0;
}
<@\textcolor{blue}{for}@> (i = 0; i < nm; i++)
{
    scanf("%d" , &nm);
    c = n[i];
}
<@\textcolor{blue}{return}@> 0;
}  
            
\end{lstlisting}
\end{minipage}
\captionof{figure}{Case 2 indicating the error in assignment statement between variable and array identifier}
\noindent
\\
\textbf{Case 3:} If the non-terminal of \texttt{rvalue} and non-terminal of \texttt{lvalue} are the children nodes of the non-terminal \texttt{BinaryOp}, then the type of \texttt{rvalue} is assigned as the type of \texttt{lvalue}. In the Figure 9 on the left side of the figure, in the conditional expression statement \textbf{$count > max$}, \texttt{BinaryOp} is "$>$", the \texttt{lvalue} "count" is undeclared with its non-terminal being \texttt{ID} and the \texttt{rvalue} is max with its non-terminal being \texttt{ID}. 
\\
\\
\begin{minipage}{.4\textwidth}
\begin{lstlisting}[xleftmargin=.2\textwidth, linebackgroundcolor={%
        \ifnum\value{lstnumber}=18
                \color{red!40}
            \fi},linebackgroundwidth=15em,numbersep=10pt,basicstyle=\small]
<@\textcolor{blue}{int}@> main()
{
<@\textcolor{blue}{int}@> n,i,j,max;
<@\textcolor{blue}{int}@> a[20];
<@\textcolor{blue}{for}@>(i=0;i<n;i++)
{
    <@\textcolor{blue}{for}@>(j=i;j<n;j++)
    {
        <@\textcolor{blue}{if}@>(a[i]<a[j])
        {
        count=count+1;
        }
    }
<@\textcolor{blue}{if}@>(count>max){max=count;}
        
}
printf("%d",max);
<@\textcolor{blue}{return}@> 0;
}            
\end{lstlisting}
\end{minipage}\hfill\vline\hfill
\begin{minipage}{.8\textwidth}
\begin{lstlisting}[xrightmargin=.5\textwidth,linebackgroundcolor={%
        \ifnum\value{lstnumber}=3
                \color{green!40}
    \fi},linebackgroundwidth=14.5em,numbersep=5pt,basicstyle=\small]
<@\textcolor{blue}{int}@> main()
{
<@\textcolor{blue}{int}@> count;
<@\textcolor{blue}{int}@> n;
<@\textcolor{blue}{int}@> i;
<@\textcolor{blue}{int}@> j;
<@\textcolor{blue}{int}@> max;
<@\textcolor{blue}{int}@> a[20];
<@\textcolor{blue}{for}@> (i = 0; i < n; i++)
{
  <@\textcolor{blue}{for}@> (j = i; j < n; j++)
  {
    <@\textcolor{blue}{if}@> (a[i] < a[j])
    {
      count = count + 1;
    }
  }
  <@\textcolor{blue}{if}@> (count > max)
  {
    max = count;
  }
}
printf("%d", max);
<@\textcolor{blue}{return}@> 0;
}
\end{lstlisting}
\end{minipage}
\captionof{figure}{Case 3 illustrating the repair of assignment statement of variable and array identifier}
\noindent
\\
\textbf{Case 4:} This case is similar to case 2 but deals with assignment of a variable to another variable instead of array element. In this Figure 10, in the left side of the figure, the \texttt{lvalue} variable "z" inside the \texttt{For} statement is undeclared, and its non-terminal node is \texttt{ID}, it is assigned to the \texttt{integer} type of the variable "i" whose non-terminal is \texttt{ID}.
\\
\\
\begin{minipage}{.4\textwidth}
\begin{lstlisting}[xleftmargin=.2\textwidth, linebackgroundcolor={%
        \ifnum\value{lstnumber}=6
                \color{red!40}
            \fi},linebackgroundwidth=15em,numbersep=10pt,basicstyle=\small]
<@\textcolor{blue}{int}@> main() {
<@\textcolor{blue}{int}@> n, i, j, k;
scanf("%d", &n); 
<@\textcolor{blue}{for}@>(i=1; i<=n; i++)
{
<@\textcolor{blue}{for}@>(j=1,z=i;j<=i;j++,k--)
    {
        <@\textcolor{blue}{if}@>((k%2) == 0)
         printf("*");
    } 
}
<@\textcolor{blue}{return}@> 0;
}            
\end{lstlisting}
\end{minipage}\hfill\vline\hfill
\begin{minipage}{.8\textwidth}
\begin{lstlisting}[xrightmargin=.5\textwidth, linebackgroundcolor={%
        \ifnum\value{lstnumber}=3
                \color{green!40}
    \fi},linebackgroundwidth=17.5em,numbersep=10pt,basicstyle=\small]
<@\textcolor{blue}{int}@> main()
{
<@\textcolor{blue}{int}@> z;
<@\textcolor{blue}{int}@> n;
<@\textcolor{blue}{int}@> i;
<@\textcolor{blue}{int}@> j;
<@\textcolor{blue}{int}@> k;
scanf("%d", &n);
<@\textcolor{blue}{for}@> (i = 1; i <= n; i++)
{
<@\textcolor{blue}{for}@> (j = 1,z = i;j<=i;j++,k--)
  {
    <@\textcolor{blue}{if}@> ((k % 2) == 0)
      printf("*");
  }
}
<@\textcolor{blue}{return}@> 0;
}
\end{lstlisting}
\end{minipage}    
\captionof{figure}{Case 4 illustrating the fix of variable "z" from the for loop statement}
\noindent
\\
\textbf{Case 5:} This case deals with binary operation involved in an assignment expression statement. In Figure 11, the terminal variable "t" is undeclared and is assigned to the type of the terminal variable "summation" which is of type \texttt{double}. In the statement \textbf{summation = summation + t*delx}, there is non-terminal \texttt{Assignment} and its children nodes being the non-terminal \texttt{ID} with terminal "summation" variable and the node \texttt{BinaryOp} with its corresponding children nodes \texttt{ID}:summation, \texttt{BinaryOp:+}, \texttt{BinaryOp:*} with its children \texttt{ID}:t, \texttt{ID}:delx. 
\\
\\
\begin{minipage}{.4\textwidth}
\begin{lstlisting}[xleftmargin=.2\textwidth, linebackgroundcolor={%
        \ifnum\value{lstnumber}=8
                \color{red!40}
            \fi},linebackgroundwidth=17em,numbersep=5pt,basicstyle=\small]
<@\textcolor{blue}{double}@> sum(<@\textcolor{blue}{double}@> a, <@\textcolor{blue}{double}@> n, <@\textcolor{blue}{double}@> delx)
{
<@\textcolor{blue}{double}@> summation=0;
<@\textcolor{blue}{int}@> j;
<@\textcolor{blue}{for}@> (j=0;j<n;j++)
{<@\textcolor{blue}{double}@> x=a+j*delx;
<@\textcolor{blue}{double}@> r=fabs(f(x)-g(x));
summation=summation+t*delx;
}
<@\textcolor{blue}{return}@> summation;
}            
\end{lstlisting}
\end{minipage}\hfill\vline\hfill
\begin{minipage}{.8\textwidth}
\begin{lstlisting}[xrightmargin=.5\textwidth, linebackgroundcolor={%
        \ifnum\value{lstnumber}=3
                \color{green!40}
    \fi},linebackgroundwidth=19em,numbersep=5pt,basicstyle=\small]
<@\textcolor{blue}{double}@> sum(<@\textcolor{blue}{double}@> a, <@\textcolor{blue}{double}@> n, <@\textcolor{blue}{double}@> delx)
{
<@\textcolor{blue}{double}@> t;
<@\textcolor{blue}{double}@> summation = 0;
<@\textcolor{blue}{int}@> j;
<@\textcolor{blue}{for}@> (j = 0; j < n; j++)
{
<@\textcolor{blue}{double}@> x = a + (j * delx);
<@\textcolor{blue}{double}@> r =fabs(f(x)-g(x));
summation =summation+(t * delx);
}

<@\textcolor{blue}{return}@> summation;
}    
\end{lstlisting}
\end{minipage}              
\captionof{figure}{Case 5 demonstrating the error in binary operation and undeclared "t" getting fixed}
\noindent
\\
\textbf{Case 6:} This case is an exact opposite of case 2 where the \texttt{lvalue} in the Figure 12 is an array identifier "b" undeclared, whose non-terminal node is \texttt{ID} and its parent node is \texttt{ArrayRef} and \texttt{rvalue} is a terminal variable "count" whose non-terminal node \texttt{ID} is assigned as \texttt{integer}. 
\\
\\
\begin{minipage}{.4\textwidth}
\begin{lstlisting}[xleftmargin=.2\textwidth, linebackgroundcolor={%
        \ifnum\value{lstnumber}=15
                \color{red!40}
            \fi},linebackgroundwidth=15em,numbersep=10pt,basicstyle=\small]
<@\textcolor{blue}{int}@> main()
{
<@\textcolor{blue}{int}@> i,j,n,k,count=0,max;
scanf("%d",&n);
<@\textcolor{blue}{int}@> a[n];
<@\textcolor{blue}{for}@>(i=0;i<n;i++){
   scanf("%d",a[i]);
}
<@\textcolor{blue}{for}@> (i=0;i<n;i++){
    <@\textcolor{blue}{for}@> (j=i;j<n;j++){
        <@\textcolor{blue}{if}@> (a[j]>a[i]){
            count++;
        }
    }
    b[i]=count;
    count=0;
}
<@\textcolor{blue}{return}@> 0;
}            
\end{lstlisting}
\end{minipage}\hfill\vline\hfill
\begin{minipage}{.8\textwidth}
\begin{lstlisting}[xrightmargin=.5\textwidth, linebackgroundcolor={%
        \ifnum\value{lstnumber}=3
                \color{green!40}
    \fi},linebackgroundwidth=14.5em,numbersep=10pt,basicstyle=\small]
<@\textcolor{blue}{int}@> main()
{
<@\textcolor{blue}{int}@> b[1000];
<@\textcolor{blue}{int}@> i;
<@\textcolor{blue}{int}@> j;
<@\textcolor{blue}{int}@> n;
<@\textcolor{blue}{int}@> k;
<@\textcolor{blue}{int}@> count = 0;
<@\textcolor{blue}{int}@> max;
scanf("%d", &n);
<@\textcolor{blue}{int}@> a[n];
<@\textcolor{blue}{for}@> (i = 0; i < n; i++)
{
  scanf("%d", a[i]);
}
<@\textcolor{blue}{for}@> (i = 0; i < n; i++)
{
  <@\textcolor{blue}{for}@> (j = i; j < n; j++)
  {
    <@\textcolor{blue}{if}@> (a[j] > a[i])
    {
      count++;
    }
  }
  b[i] = count;
  count = 0;
}
<@\textcolor{blue}{return}@> 0;
}
\end{lstlisting}
\end{minipage}               
\captionof{figure}{Illustration of case 6 marked by red line indicating the error in assignment statement}
\noindent
\\
\textbf{Case 7:} This case is slightly similar to case 5 but it does not involve any assignment operation. The type can be assigned to a variable not only from \texttt{lvalue} but also from its neighbouring variables involved in a binary operation. As we can see in left side of the Figure 13, the terminal variable "diff" being undeclared is involved in a binary operation \texttt{BinaryOp:*} and \texttt{BinaryOp:+} with the terminal variables "key" and "a" respectively, so the variable "diff" is assigned to \texttt{integer} type.
\\
\\
\begin{minipage}{.4\textwidth}
\begin{lstlisting}[xleftmargin=.2\textwidth, linebackgroundcolor={%
        \ifnum\value{lstnumber}=8
                \color{red!40}
            \fi},linebackgroundwidth=16em,numbersep=5pt,basicstyle=\small]
<@\textcolor{blue}{int}@> main()
{
<@\textcolor{blue}{const double}@> E=0.000001;
<@\textcolor{blue}{double}@> a,b,inter,subarea=0;
<@\textcolor{blue}{int}@> n,key=0;
scanf("%lf%lf%d",&a,&b,&n);
inter=(b - a)/n;
<@\textcolor{blue}{while}@>(key<n&&diff*key+1< E)
{
    subarea+=1;
    key++;
}
<@\textcolor{blue}{return}@> 0;
}            
\end{lstlisting}
\end{minipage}\hfill\vline\hfill
\begin{minipage}{.8\textwidth}
\begin{lstlisting}[xrightmargin=.5\textwidth, linebackgroundcolor={%
        \ifnum\value{lstnumber}=3
                \color{green!40}
    \fi},linebackgroundwidth=20em,numbersep=5pt,basicstyle=\small]
<@\textcolor{blue}{int}@> main()
{
<@\textcolor{blue}{int}@> diff;
<@\textcolor{blue}{const double}@> E = 0.000001;
<@\textcolor{blue}{double}@> a;
<@\textcolor{blue}{double}@> b;
<@\textcolor{blue}{double}@> inter;
<@\textcolor{blue}{double}@> subarea = 0;
<@\textcolor{blue}{int}@> n;
<@\textcolor{blue}{int}@> key = 0;
scanf("%lf""%lf""%d", &a, &b, &n);
inter = (b - a) / n;
<@\textcolor{blue}{while}@>((key < n)&&((diff*key)+ 1)<E))
{
  subarea += 1;
  key++;
}
<@\textcolor{blue}{return}@> 0;
}
\end{lstlisting}
\end{minipage}            
\captionof{figure}{Demo of case 7 involving the error in while loop statement}
\noindent
\\
\textbf{Case 8:} In this case, the type of a variable is bound from the type of a function call in a conditional expression. In Figure 14, the terminal variable "k" is undeclared in the "for" expression \textbf{k $>=$ hanoi(j)-1} and it is being involved in a binary operation \texttt{BinaryOp:>=} with the function call \textbf{hanoi(j)} where the corresponding non-terminal node of the terminal "hanoi" is \texttt{ID} and parent node being \texttt{FuncCall}, is \texttt{integer}. 
\\ 
\\
\begin{minipage}{.4\textwidth}
\begin{lstlisting}[xleftmargin=.2\textwidth, linebackgroundcolor={%
        \ifnum\value{lstnumber}=8
                \color{red!40}
            \fi},linebackgroundwidth=16em,numbersep=5pt,basicstyle=\small]
<@\textcolor{blue}{int}@> main() {
<@\textcolor{blue}{int}@> t,i,n,j;
<@\textcolor{blue}{int}@> x;
scanf("%d",&t);
<@\textcolor{blue}{for}@>(i=1;i<t;i++)
{  
scanf("%d",&n);
<@\textcolor{blue}{for}@>(j=0;k>=hanoi(j)-1;j++)
{
    <@\textcolor{blue}{if}@>(hanoi(j)-1==k)
        printf("yes");
    <@\textcolor{blue}{else}@>
        printf("no");
}
}
<@\textcolor{blue}{return}@> 0;
}            
\end{lstlisting}
\end{minipage}\hfill\vline\hfill
\begin{minipage}{.8\textwidth}
\begin{lstlisting}[xrightmargin=.5\textwidth, linebackgroundcolor={%
        \ifnum\value{lstnumber}=3
                \color{green!40}
    \fi},linebackgroundwidth=20.5em,numbersep=5pt,basicstyle=\small]
<@\textcolor{blue}{int}@> main()
{
<@\textcolor{blue}{int}@> k;
<@\textcolor{blue}{int}@> t;
<@\textcolor{blue}{int}@> i;
<@\textcolor{blue}{int}@> n;
<@\textcolor{blue}{int}@> j;
<@\textcolor{blue}{int}@> x;
scanf("%d", &t);
<@\textcolor{blue}{for}@> (i = 1; i < t; i++)
{
  scanf("%d", &n);
  <@\textcolor{blue}{for}@> (j = 0;k>=(hanoi(j) - 1);j++)
  {
  <@\textcolor{blue}{if}@> ((hanoi(j) - 1) == k)
    printf("yes");
  <@\textcolor{blue}{else}@>
    printf("no");
  }
}
<@\textcolor{blue}{return}@> 0;
}
\end{lstlisting}
\end{minipage} 
\captionof{figure}{Case 8 indicating the undeclared "k" in for loop statement}
\noindent
\\
\textbf{Case 9:} This case is similar to case 8, however instead of a conditional expression with a binary operation, the type of the function call is a \texttt{rvalue} is bound to a \texttt{lvalue} variable in an assignment expression statement with a binary operation. This can be seen from Figure 15, where variable "y" is undeclared in the assignment expression \textbf{y = tower(j)-1} and is assigned to the type of the function call \textbf{tower(j)} whose non-terminal node is \texttt{ID} and parent node is \texttt{FuncCall}.
\\
\\
\begin{minipage}{.4\textwidth}
\begin{lstlisting}[xleftmargin=.2\textwidth, linebackgroundcolor={%
        \ifnum\value{lstnumber}=9
                \color{red!40}
            \fi},linebackgroundwidth=15em,numbersep=10pt,basicstyle=\small]
<@\textcolor{blue}{int}@> main() 
{<@\textcolor{blue}{int}@> i,n,j,t;
 scanf("%d\n",&n);
 <@\textcolor{blue}{for}@>(i=1;i<=n;i++)
 {
   scanf("%d\n",&t);
 <@\textcolor{blue}{for}@>(j=1;j<=200;j++)
 {
  y=tower(j)-1;
 }
 }
<@\textcolor{blue}{return}@> 0;
}            
\end{lstlisting}
\end{minipage}\hfill\vline\hfill
\begin{minipage}{.8\textwidth}
\begin{lstlisting}[xrightmargin=.5\textwidth, linebackgroundcolor={%
        \ifnum\value{lstnumber}=3
                \color{green!40}
    \fi},linebackgroundwidth=16.5em,numbersep=10pt,basicstyle=\small]
<@\textcolor{blue}{int}@> main()
{
<@\textcolor{blue}{int}@> y;
<@\textcolor{blue}{int}@> i;
<@\textcolor{blue}{int}@> n;
<@\textcolor{blue}{int}@> j;
<@\textcolor{blue}{int}@> t;
scanf("%d""\n", &n);
<@\textcolor{blue}{for}@> (i = 1; i <= n; i++)
{
  scanf("%d""\n", &t);
  <@\textcolor{blue}{for}@> (j = 1; j <= 200; j++)
  {
    y = tower(j) - 1;
  }
}
<@\textcolor{blue}{return}@> 0;
}
\end{lstlisting}
\end{minipage}            
\captionof{figure}{Case 9 demonstrating the undeclared identifier "y" in assignment statement with a function call}
\begin{table*}
\begin{tabular}{ | P{2cm} | P{1.7cm} | P{1.5cm}| P{1.7cm}| P{1.5cm}| P{1.7cm}| P{1.7cm}| P{1.2cm}| }
 \cline{2-8}
 \multicolumn{1}{c|}{} & \textbf{Identified} & \textbf{Not Identified} & \textbf{Correctly Identified (True Positive)} & \textbf{Wrongly Identified (False Positive)} & \textbf{Correctly Identified + Correct Type Inferred (Fixed)} & \textbf{Wrongly Identified + Wrong Type Inferred (Not Fixed)} & \textbf{Total} \\
 \hline
 \textbf{Undeclared Variables and Arrays} & 887(83.7\%) & 172 & 857(80.9\%) & 202 & 844(79.7\%) & 215 & 1059 \\
 \hline
 \textbf{Undeclared variables - Main function} & N/A & N/A & 566(99.1\%) & 5 & 560(98\%) & 11 & 571   \\
\hline
 \textbf{Undeclared variables - Multiple functions} & N/A & N/A & 179(91.7\%) & 16 & 172(88.2\%) & 23 & 195 \\
\hline
 \textbf{Undeclared Arrays - Main functions} & N/A & N/A &  90(96.8\%) & 3 & 90(96.8\%) & 3 & 93 \\
\hline
 \textbf{Undeclared Arrays - Multiple functions} & N/A & N/A &  22(78.5\%) & 6 & 22(78.5\%) & 6 & 28 \\
 \hline
\end{tabular}
\caption{\textbf{Analysis results of both the undeclared variables and arrays}}
\end{table*}
\noindent
\\
Table 1 shows the results of analysis obtained after performing the compilation of the programs manually where the first row consists of all the programs (1059) that are containing both undeclared variables and arrays in which our approach has located and identified the undeclared variables in 887(83.7\%) programs out of total number of 1059 programs. However, our approach has correctly located and identified them in 857(80.9\%) programs, but the repair is performed on 844(79.7\%) programs by correctly locating as well as inferring and binding their types. The value of the first column "Identified" in the rest of the rows are not applicable (N/A) as the results correspond to programs where undeclared variables are identified and located. Similarly, the second row shows the results for programs with only undeclared variables and consisting of only single main functions (571) out of the identified programs (887). The third row displays the results of programs with undeclared variables and containing two or more functions including main function (195) out of 887 programs. The fourth row demonstrates the results shown by programs only with errors caused due to undeclared arrays and having one and only main function (93) out of the 887 programs. Finally, the last row illustrates the number of programs which contains undeclared variables and having two or more functions along with main function (22) out of those 887 programs. Table 2 shows the summary of various cases along the rows and its brief description message along the columns through which the type binding is performed before the compile-time.
\noindent
\\
\\
\begin{table*}
\begin{tabular}{ | p{1.6cm} | p{14cm} | }
 \hline
 \textbf{Cases} & \textbf{Brief Description} \\
 \hline
 Case 1 & Assignment expression statement with a constant on the right-hand side of the expression \\
 \hline
 Case 2  & Assignment expression statement with an array identifier on the right-hand side and an identifier other than array identifier on the left-hand side of the expression  \\
\hline
 Case 3 & Conditional expression statement with a binary operation between the identifier variables \\
 \hline
 Case 4 & Assignment expression statement with an identifier other than array identifier on the right-hand side of the expression \\
 \hline
 Case 5 & Assignment expression statement with a binary operation between the identifier variables \\
 \hline
 Case 6 & Assignment expression statement with an  identifier other than arrays on the right-hand side and an array identifier on the left-hand side of the expression \\
 \hline
 Case 7 & Binary operation between identifier variables in a loop expression statement \\
 \hline
 Case 8 & Conditional expression statement with a binary operation between an identifier and a function call expression \\
 \hline
 Case 9 & Assignment expression statement with a binary operation between an identifier and a function call expression\\
 \hline
\end{tabular}
\caption{\textbf{Summary of Type Binding Case Description}}
\end{table*}
\begin{minipage}{.4\textwidth}
\begin{lstlisting}[xleftmargin=.1\textwidth, frame=tlrb,linebackgroundcolor={%
        \ifnum\value{lstnumber}=3
                \color{green!40}
            \fi},linebackgroundwidth=15em,numbersep=8pt,basicstyle=\small]
<@\textcolor{blue}{int}@> main()
{
<@\textcolor{blue}{int}@> J;
<@\textcolor{blue}{int}@> n;
<@\textcolor{blue}{int}@> i;
<@\textcolor{blue}{int}@> j;
<@\textcolor{blue}{int}@> flag = 0;
scanf("%d", &n);
<@\textcolor{blue}{int}@> a[51];
<@\textcolor{blue}{for}@> (i = 0; i < n; i++)
{
  scanf("%d", &a[i]);
}
<@\textcolor{blue}{for}@> (i = 0; i < n; i++)
{
  <@\textcolor{blue}{for}@> (j = 0; j < n; J++)
  {
    <@\textcolor{blue}{if}@> (a[i] == a[j])
    {
      printf("YES");
      flag = 1;
      break;
    }
  }
}
<@\textcolor{blue}{return}@> 0;
}
\end{lstlisting}
\end{minipage}\hfill
\begin{minipage}{.5\textwidth}
\begin{lstlisting}[frame=tlrb,linebackgroundcolor={%
        \ifnum\value{lstnumber}=3
                \color{green!40}
    \fi},linebackgroundwidth=14.5em,numbersep=8pt,basicstyle=\small]
<@\textcolor{blue}{int}@> main()
{
<@\textcolor{blue}{int}@> l;
<@\textcolor{blue}{double}@> a;
<@\textcolor{blue}{double}@> b;
<@\textcolor{blue}{double}@> k;
<@\textcolor{blue}{double}@> p;
<@\textcolor{blue}{int}@> n;
scanf("%f" "%f" "%d", &a, &b, &n);
k = ((a - b) * 1.0) / n;
<@\textcolor{blue}{for}@> (l = 1; l <= n; l++)
{
  <@\textcolor{blue}{if}@>((l * k)<(-1))
    p+= k;

  <@\textcolor{blue}{if}@>((l*k>=-1) && (l*k<= 1))
    p=p + (((l * k)*(l * k))*k);

  <@\textcolor{blue}{if}@>((l * k) > 1)
    p=p + ((l * k)*(l * k))*(l*k)*k;
  }
printf("%.4f", p);
<@\textcolor{blue}{return}@> 0;
}    
\end{lstlisting}
\end{minipage}   
\captionof{figure}{Picture on left illustrates repair that caused infinite loop due to variable "J" incremented in loop statement and the right side depicts repair caused by binding type \texttt{int} instead of \texttt{double}}
\section{Discussion}
There are few limitations in our approach. Fixing the undeclared variables that had been due to imperceptible spelling mistakes or a variable that is used only once throughout the program may cause the program to run in an infinite loop or lead to some possible run-time errors. As seen in the left side of Figure 16, instead of incrementing the variable "j" inside the \texttt{for} loop, the programmer had used "J" instead which had caused the program to run into an infinite loop. Another major limitation is in the type binding approach on the right side of Figure 16 shows an example where, the type has been wrongly bound to the variable "l" because "l" is used in the \texttt{for} loop expression in which "l" is assigned to constant "1"  as well as it is used in an assignment expression inside the \texttt{for} loop body, in the statement \textbf{if((l*k) $<$ -1)}, variable "k" is of the type \texttt{double} and "*" is a binary operation, so the variable "j" should be assigned of the type \texttt{double} instead it is inferred as an \texttt{integer} type due to the former case.
\\
\\
We had seen in our model that a vocabulary in the form of hash table is used for training purposes. The purpose of training neural networks on the hash table is due to the fact that it can be used for recognizing input patterns (keys) in the hash table and can be used to predict the sequences (values). Consider the case where an input pattern is not present in the hash table and we need to predict the sequence, a hash table would have return null in this case but neural networks will give the closest sequence prediction. 
\\
\\
The benefits of our approach lies in the fact that our model could be used in real-time as a tool for any C programming environment online or offline editors in locating, reporting and repairing undeclared identifiers for any C programs. Additionally, our model can be used when there are lack of positive bug-free syntactically correct and executing source program reference examples for buggy source programs. Also, our type binding approach will be applicable even for declared variables.
\\
\section{Conclusion and Future Work}
In this paper, we had seen different cases of one of the most common semantic error: undeclared variables. We had combined AST and LSTM approaches to extract a set of non-terminal and terminal nodes to carry out the classification and prediction tasks of the undeclared variables. We had also seen the generation of clean and buggy-free source programs by performing AST transformation and serialization as well as deserialization of AST to JSON and vice- versa. Furthermore, in this paper we had coined a new term known as Pre-Compile Time Type binding where we had implemented the fix of the types of undeclared variables by binding them their corresponding types before providing it for the compiler to compile them. By our approach, we had correctly identified 81$\%$ of the programs that contains only undeclared identifier errors. Also, we had fixed those undeclared identifier errors by binding their corresponding types in 80$\%$ of the programs. 
\\
\\
In future, we would like to perform automatic repair on different types of syntactic,semantic errors and logical errors. Further, we plan to perform type binding for the limitation cases in Figure 16 as well as also implement a repair approach for the logical errors that arises after the repair of syntactic and semantic errors caused by variables used only once in the program or due to spelling mistakes.
\bibliographystyle{unsrt}  


\end{document}